\documentclass[a4paper,11pt]{article}

\usepackage{amsmath, amsfonts, amssymb}
\usepackage{graphicx}
\usepackage{float}
\usepackage{natbib}
\usepackage{geometry}
\usepackage{subcaption}
\usepackage{enumerate}
\usepackage{orcidlink}

\usepackage{dcolumn,booktabs}
\newcolumntype{d}[1]{D{.}{.}{#1}}

\usepackage[symbol]{footmisc}

\usepackage{authblk}

\usepackage{hyperref}

\newcommand{\R}{\textsf{R}}

\newcommand{\vect}[1]{\boldsymbol{#1}}

\newcommand{\keywords}[1]{
    \noindent
    \textbf{Keywords:} #1
}

\title{Bayesian mortality forecasting with a Conway--Maxwell--Poisson specification}

\author[a]{Jackie Siaw Tze Wong\, \orcidlink{0000-0002-0314-6684}}
\author[b]{Emiliano A. Valdez, \orcidlink{0000-0003-4037-8385}}

\affil[a]{\small
School of Mathematics, Statistics, and Actuarial Science, University of Essex,  
Wivenhoe Park, Colchester, CO4 3SQ, UK. 
Email: \href{mailto:jw19203@essex.ac.uk}{\texttt{jw19203@essex.ac.uk}}.  
\emph{Corresponding author.}
}

\affil[b]{\small
Department of Mathematics, University of Connecticut,  
Storrs, CT 06269-1009, USA.
Email: \href{mailto:emiliano.valdez@uconn.edu}{\texttt{emiliano.valdez@uconn.edu}}.
}

\begin{document}

\maketitle

\begin{abstract} 

This paper presents a novel approach to stochastic mortality modelling by using the Conway--Maxwell--Poisson (CMP) distribution to model death counts. Unlike standard Poisson or negative binomial distributions, the CMP is a more adaptable choice because it can account for different levels of variability in the data, a feature known as dispersion. Specifically, it can handle data that are underdispersed (less variable than expected), equidispersed (as variable as expected), and overdispersed (more variable than expected). We develop a Bayesian formulation that treats the dispersion level as an unknown parameter, using a Gamma prior to enable a robust and coherent integration of the parameter, process, and distributional uncertainty. The model is calibrated using Markov chain Monte Carlo (MCMC) methods, with model performance evaluated using standard statistical criteria such as residual analysis and scoring rules. An empirical study using England and Wales male mortality data shows that our CMP-based models provide a better fit for both existing data and future predictions compared to traditional Poisson and negative binomial models, particularly when the data exhibit overdispersion. Finally, we conduct a sensitivity analysis with respect to prior specification to assess robustness. 

\end{abstract}

\vspace{0.8cm}


\pagebreak

\section{Introduction}
\label{sec:intro} 
 
Mortality forecasting is vital for demographers, actuaries, and policymakers because it provides insights into future population dynamics, the sustainability of retirement and social security systems, and the solvency of insurance companies. A major research advance came with the Lee--Carter (LC) model (\citealp{lca}), which introduced a simple yet powerful stochastic framework to decompose and extrapolate age-specific death rates. \cite{basellini2023} and \cite{booth2008} provide comprehensive surveys of subsequent developments. While foundational models such as LC often assume that death counts follow simple distributions, like the Poisson, an important area of modern research focuses on properly modelling dispersion, the unexpected noise in the death data that the main model structure does not typically capture. It is crucial to account for overdispersion, and failure to accurately model this extra variability results in model's prediction intervals that are too narrow. This, in turn, leads to an underestimation of longevity risk and may result in significant financial consequences.

In this paper, we present three main contributions. First, we introduce the use of Conway--Maxwell--Poisson (CMP) distribution as a flexible approach for modelling death counts in the presence of dispersion. Originally proposed in \cite{Conway_1962}, the CMP distribution has not, to the best of our knowledge, previously been applied in the context of mortality modelling. This model allows us to capture under-, equi-, and over-dispersion in mortality data and provides us an opportunity to illustrate its versatility and practical relevance for demographic and actuarial applications.

Secondly, we develop a Bayesian formulation of the proposed model in which the dispersion is treated as an unknown but estimable parameter and assigned a Gamma prior. This specification enables explicit inference on both the nature and level of dispersion, while coherently integrating multiple sources of uncertainty within a unified probabilistic framework. The advantages of the Bayesian approach are well documented in the literature, such as \cite{cairns2011} and \cite{wong_2017}. In addition, there is an emerging body of recent work applying Bayesian methods in mortality and longevity modelling, e.g. \cite{Diana_Wong_Pittea_2025}, \cite{goes2025}, \cite{shi2024multi}.

Finally, we demonstrate the proposed methodology using England and Wales male death data. The empirical results are encouraging and show clear improvements over the conventional Poisson and negative binomial specifications.

The rest of the paper is organised as follows. The remainder of Section \ref{sec:intro} introduces the data, the notation, and the Poisson specification for modelling death counts. Section \ref{sec:CMP} presents the CMP distribution as a flexible alternative to handle overdispersion. Section \ref{sec:bayesian} outlines the Bayesian implementation, including the prior specification and the corresponding estimation algorithm. Empirical results based on England and Wales male mortality data are reported in Section \ref{sec:results}. Finally, we conclude with a sensitivity analysis on the prior specification and some closing remarks.

\subsection{Data and notation}
\label{sec:data}

Let $d_{xt}$ denote the observed number of deaths for age group $x$ in year $t$, where $x=\{x_1,\ldots,x_A\}$ and $t=\{t_1,\ldots,t_T\}$ represent a set of $A$ different age groups and $T$ different years, respectively. Suppose further that $e_{xt}$ and $\mu_{xt}$ respectively denote the corresponding central exposure to risk and central mortality rate.

The data chosen for illustrative purposes are male death data and the corresponding exposures of England and Wales, extracted from the Human Mortality Database (HMD)\footnote[2]{ See http://www.mortality.org.}. They are classified by single year of age from $0$ to $99$, and years ranging from $1961$ to $2002$. Hence, here we have $\{x_1,\ldots,x_A\}=\{0,\ldots,99\}$ and $\{t_1,\ldots,t_T\}=\{1961,\ldots,2002\}$ with $A=100$ and $T=42$. The data for years 2003-2021 were reserved as a validation set (see Section \ref{sec:results}).

\subsection{Poisson distribution for modelling death counts}
\label{sec:plc}

The Poisson distribution is commonly employed to model death counts (see, for example, \citealp{mccullagh2019generalized} or \citealp{selvin2004statistical}). More recent applications include \cite{hayat2014understanding}, \cite{zero_inflate_poisson}, \cite{covid_poisson_arima} etc. Let $D_{xt}$ be the random variable denoting the number of deaths of age $x$ year $t$, then
\begin{eqnarray}
D_{xt} \vert \mu_{xt} \sim {\rm Poisson}(e_{xt}\mu_{xt}), 
\label{eqn:poisson_Dxt}
\end{eqnarray}
with the corresponding probability mass function (PMF) 
\begin{eqnarray}
\mathbb{P}[D_{xt}=d_{xt} \vert \mu_{xt}] = \frac{(e_{xt}\mu_{xt})^{d_{xt}}\exp(-e_{xt}\mu_{xt})}{d_{xt}!} \propto \mu_{xt}^{d_{xt}}\exp(-e_{xt}\mu_{xt}).
\label{eqn:poisson_pmf_Dxt}
\end{eqnarray}

There are several structural assumptions for $\mu_{xt}$, each defining different mortality rate models in the literature. For example, specifying the popular Lee--Carter model (\citealp{lca}) gives:
\begin{eqnarray}
\left\{\begin{array}{l}
D_{xt}\vert \mu_{xt} \sim {\rm Poisson}(e_{xt}\mu_{xt})\ , \\
\log\mu_{xt} = \alpha_x+\beta_x\kappa_t \ ,
\end{array}\right.
\label{eqn:plc}
\end{eqnarray}
where the following constraints are adopted for model identifiability,
$$\sum_{x}\beta_x=1 \mbox{ \ \  and  \ \ } \sum_{t}\kappa_t=0.$$ 
This yields the \textbf{Poisson Lee--Carter (Poisson--LC)} as proposed by \cite{poissonlca}. The parameters can be interpreted as follows: 
\begin{eqnarray*}
\alpha_x &:& \mbox{ is the average of the logarithm of death rates over time (i.e.,\ $\alpha_{x}=\frac{\sum_{t}\log\mu_{xt}}{T}$).} \\
\beta_x &:& \mbox{ is the age-specific pattern of mortality improvement, quantifying the sensitivity of} \\
&& \mbox{ mortality at each age to overall changes.} \\
\kappa_t &:& \mbox{ represents the overall temporal trend in mortality, appropriately modulated by $\beta_x$.}
\end{eqnarray*}

In this paper, we also consider the inclusion of cohort effects for comparison purposes. Following the formulation of \cite{RH_cohort}, 
\begin{eqnarray}
\left\{\begin{array}{l}
D_{xt}\vert \mu_{xt}\sim {\rm Poisson}(e_{xt}\mu_{xt})\ , \\
\log\mu_{xt} = \alpha_x+\beta_x\kappa_t + \gamma_c\ ,
\end{array}\right.
\label{eqn:plcc}
\end{eqnarray}
where $\gamma_c$ are cohort parameters with $c=t-x\in\{1,\ldots,C\}$ denoting the cohort index, which represent cohorts born in $\{1861,\ldots,2001\}$, and $C=A+T-1$. We will follow the suggestion of \cite{wong_2023} in imposing the following set of constraints\footnote[2]{We note that the constraints are different from both \cite{RH_cohort} and \cite{hunt_2015}, but this choice leads to stable computations with good convergence properties.}:
\begin{eqnarray}
\sum_t \kappa_t=\sum_t t\kappa_t=\sum_c \gamma_c=\sum_c c\gamma_c=\sum_c c^2\gamma_c=0,
\label{eqn:apci_cons}
\end{eqnarray}
We refer to this model as the \textbf{Poisson Lee--Carter with cohorts (Poisson--LCC)} model.

\subsection{Overdispersion}
\label{sec:overdispersion}

The Poisson model imposes the mean-variance equality ($\mathbb{E}[D_{xt}]=\mathrm{Var}[D_{xt}]=e_{xt}\mu_{xt}$), a restrictive assumption that rarely holds in practice. It implies homogeneity within each age-period cell, i.e., individuals of the same age and year are assumed to share identical mortality risks. This is unrealistic, as mortality is also influenced by factors such as smoking, income, ethnicity, and genetic background \cite{brown}, leading to additional variation across individuals, commonly termed overdispersion.

\begin{figure}[H]
\centering
\captionsetup{type=figure}

\includegraphics[scale=1]{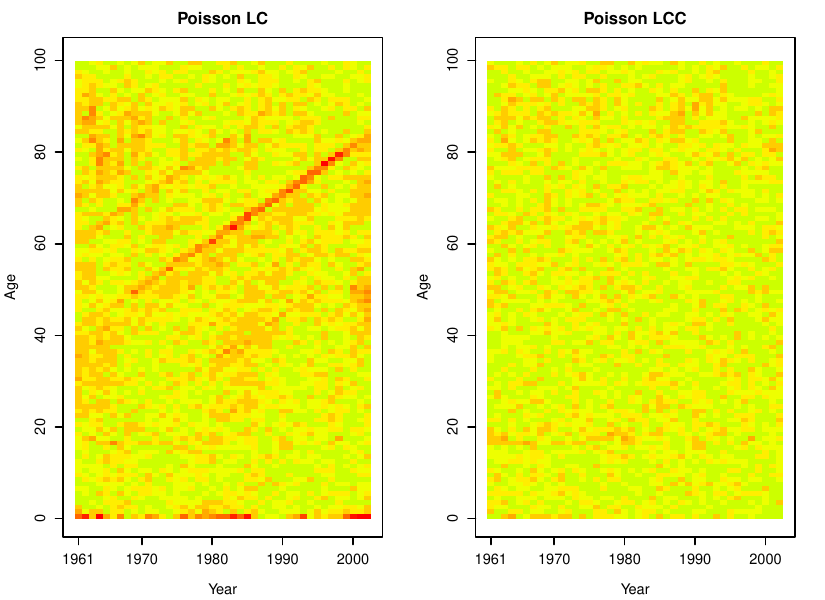}
\vspace{-0.75cm}

\captionsetup{type=figure}
\includegraphics[height=2cm,width=14cm]{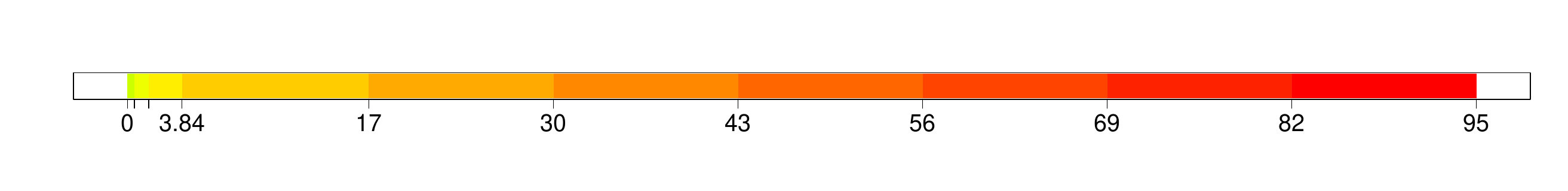}
\caption{Heat maps of $r_{xt}^2$ for the Poisson--LC (left) and Poisson--LCC (right) models. Green/yellow rectangular cells indicate areas with good fit; while orange/red-colored cells indicate areas with significantly poor fit.}
\label{fig:heatmap_plc_plcc}
\end{figure}

The issue of overdispersion due to extra-Poisson variability has been noted by several articles, e.g., \cite{brilinger_1986}, \cite{breslow_extra_poisson}, \cite{hinde_1998}, \cite{philip_lee_1997}, \cite{dean_1989}, \cite{dean_1992}. Here, we illustrate this phenomenon using the square of Pearson residuals given as:
\begin{eqnarray}
r_{xt}^2=\left.\frac{(d_{xt}-\mathbb{E}[D_{xt}])^2}{\mathrm{Var}[D_{xt}]}\right|_{\mu_{xt}=\hat{\mu}_{xt}}=\frac{(d_{xt}-e_{xt}\hat{\mu}_{xt})^2}{e_{xt}\hat{\mu}_{xt}},
\label{eqn:pearson}
\end{eqnarray}
where  $\hat{\mu}_{xt}$ is the maximum likelihood estimate (MLE) of the underlying mortality rate. Specifically,
\begin{equation*}
\hat{\mu}_{xt} =
\begin{cases}
\exp\!\left(\hat{\alpha}_x + \hat{\beta}_x \hat{\kappa}_t\right),
& \text{for Poisson--LC}, \\[4pt]
\exp\!\left(\hat{\alpha}_x + \hat{\beta}_x \hat{\kappa}_t + \hat{\gamma}_c\right),
& \text{for Poisson--LCC}.
\end{cases}
\end{equation*}
where $\hat{\alpha}_x$, $\hat{\beta}_x$, $\hat{\kappa}_t$, $\hat{\gamma}_c$ are the MLE of the parameters. A colour-coded heat map of $r_{xt}^2$ can then be constructed to visualise the lack of fit of both the Poisson--LC and Poisson--LCC models for our mortality data, as depicted in Figure \ref{fig:heatmap_plc_plcc}.

%

Under the null hypothesis that the Poisson--LC model is the true underlying model (and some mild conditions), each $r_{xt}^2$ has an approximate chi-squared distribution with degrees of freedom one ($\chi_{1}^{2}$) asymptotically. Ideally, we should expect only around $5\%$ of the rectangular cells ($AT\times 0.05=210$) to have poor fit (which we define as $r_{xt}^2>3.84$, where $3.84$ is the $95^{\rm th}$ precentile of $\chi^2_1$). However, it is evident from Figure \ref{fig:heatmap_plc_plcc} that the heat map for the Poisson--LC model is scattered with more than the expected number of orange/red cells (1128, about $27\%$), and is especially obvious for infants and cohorts born around the year 1920 (diagonal lines of orange/red cells), suggesting the model inadequacy in accounting for additional variations in the data. Even after the inclusion of cohort effects as in the Poisson--LCC model, the heat map remains scattered with orange/red cells (474, about $11\%$), indicating a poor overall fit.

Additionally, we can perform the Pearson's chi-squared overall goodness of fit test. In particular, the model deviance computed as the sum of $r_{xt}^2$,
$$r^2=\sum_{x,t}r_{xt}^2=\sum_{x,t}\frac{(d_{xt}-e_{xt}\hat{\mu}_{xt})^2}{e_{xt}\hat{\mu}_{xt}},$$  
giving values of $16709.85$ and $6628.97$ for the Poisson--LC and Poisson--LCC models respectively. Again, under the null hypothesis that this model is a good fit to the data, $r^2$ should follow an approximate chi-squared distribution with degrees of freedom ($df$) given as $df=(A-1)(T-2)=3960$ (see \citealp{RH_cohort}). Since the model deviances of $16709.85$ and $6628.97$ are substantially larger than the critical value of the conventional chi-squared statistics (i.e.\ the $95^{th}$ percentile of $\chi_{df}^{2}$ is $4107.51$), this clearly suggests that both Poisson--LC and Poisson--LCC models do not provide satisfactory fit to the data. 


As noted by \cite{wong_2017}, ignoring overdispersion leads to underestimated variance, causing overfitting and yielding overly narrow forecasts. Allowing for overdispersion produces more flexible fits, better captures heterogeneity, and generates more reliable prediction intervals (see Section \ref{sec:fitted_projected_rates}). Moreover, incorporating cohort effects further improves calibration between data and uncertainty, reducing projection biases while maintaining appropriate interval widths. In the following sections, we demonstrate how the CMP distribution addresses the extra-Poisson variation in death counts.

\section{The Conway--Maxwell--Poisson (CMP) Distribution}
\label{sec:CMP}

The CMP distribution was originally proposed by \cite{Conway_1962} as a generalization of the Poisson distribution. The PMF of a random variable $Y$ following a CMP distribution is given by:

\[
\mathbb{P}[Y=y | \lambda, \nu] = \frac{1}{Z(\lambda,\nu)}\frac{\lambda^y}{(y!)^\nu}, \ \ \ \ \ \ \text{ for } y\in\{0,1,2,\ldots\},
\]
where  
\[
Z(\lambda,\nu)=\sum_{j=0}^{\infty}\frac{\lambda^j}{(j!)^\nu}
\]
is the normalizing constant that ensures the PMF sums to 1. The model parameters satisfy $\lambda > 0$ and $\nu \geq 0$. This is commonly denoted by $Y\sim \text{CMP }(\lambda,\nu)$.

\subsection{Some properties}

The key strength of this distribution is its ability to model count data exhibiting both overdispersion and underdispersion through the dispersion parameter $\nu$. We note that the moments of $Y\sim \text{CMP }(\lambda,\nu)$ are given by
\begin{eqnarray*}
\mathbb{E}[Y] = \sum_{j=0}^{\infty} \frac{j\lambda^j}{(j!)^\nu Z(\lambda,\nu)} , \mbox{\hspace{0.5cm} and \hspace{0.5cm}}
\text{Var}[Y] = \sum_{j=0}^{\infty} \frac{j^2\lambda^j}{(j!)^\nu Z(\lambda,\nu)} - (\mathbb{E}[Y])^2.
\end{eqnarray*}

Or more specifically, when $\nu$ and $\lambda$ are not too small, we have
\begin{eqnarray}
\mathbb{E}[Y] \approx \lambda^{1/\nu} +\frac{1}{2\nu}-\frac{1}{2} \mbox{\hspace{0.5cm} and \hspace{0.5cm}}
\text{Var}[Y] \approx \frac{\lambda^1/\nu}{\nu},
\label{eqn:cmp_mean_var}
\end{eqnarray}
approximately (see for example \citealp{chanialidis2018efficient}). 
This allows for $\text{Var}[Y]> \mathbb{E}[Y]$ (overdispersion; when $\nu<1$), $\text{Var}[Y]< \mathbb{E}[Y]$ (underdispersion; when $\nu>1$) or $\text{Var}[Y]= \mathbb{E}[Y]$ (equidispersion; when $\nu=1$) depending on the value of $\nu$. 


Also note that the ratio of subsequent probabilities 
\[
\frac{\mathbb{P}[Y=y-1 | \lambda, \nu]}{\mathbb{P}[Y=y | \lambda, \nu]} = \frac{y^\nu}{\lambda},
\]
indicating that the decay rate follows one of the cases below:

\begin{itemize}
\item $\nu<1$: the rate of decay is slower than linear decrease, meaning the distribution has heavier/longer tail (overdispersion).
 \item $\nu=1$: the rate of decay is linear, which becomes the ordinary Poisson distribution (equidispersion).
 \item $\nu>1$: the rate of decay is quicker than linear decrease, meaning the distribution has lighter/shorter tail (underdispersion).
 \item $\nu=0$: this is a special case where the CMP distribution reduces to a geometric distribution with probability of $1-\lambda$ where $0<\lambda<1$.
 \item $\nu\rightarrow \infty$: this is a special case where the CMP distribution reduces to a Bernoulli distribution with probability of success $\frac{\lambda}{1+\lambda}$.
\end{itemize} 

Computations involving the CMP distribution can be carried out using the R package \emph{COMPoissonReg} (\citealp{CMPreg}). Further details on the CMP distribution are provided in \cite{daly2015conway}.

\subsection{CMP distribution for modelling death counts}
\label{sec:p_ln_lc}

This issue of overdispersion in the Poisson assumption has been explored by \cite{wong_2017} in the context of mortality modeling/forecasting, where two extensions to the above specification were proposed. In this work, we present the use of the CMP distribution as an alternative, which introduces a general dispersion parameter to relax the stringent assumptions of the Poisson model. That is, we propose
\begin{equation}
    D_{xt} \vert \mu_{xt},\nu \sim \text{CMP }\left( (e_{xt}\mu_{xt}+\frac{1}{2}-\frac{1}{2\nu})^\nu , \nu\right)\ ,
    \label{eqn:cmp_dxt}
\end{equation}
where $\nu$ is the dispersion parameter to be estimated. Suppose we write $\lambda_{xt}=(e_{xt}\mu_{xt}+\frac{1}{2}-\frac{1}{2\nu})^\nu$ for simplicity, the corresponding PMF is
\[
\mathbb{P}[D_{xt}=d_{xt} | \mu_{xt},\nu] = \frac{1}{Z(\lambda_{xt},\nu)}\frac{\lambda_{xt}^{d_{xt}}}{(d_{xt}!)^\nu}\ , 
\]
where  
\[
Z(\lambda_{xt},\nu)=\sum_{j=0}^{\infty}\frac{\lambda_{xt}^j}{(j!)^\nu}\ .
\]
Note that this (somewhat unconventional) formulation is designed to ensure that the expected number of deaths coincides with the product of exposures and mortality rates, i.e.
\begin{eqnarray}
\mathbb{E}[D_{xt} ] &\approx& ((e_{xt}\mu_{xt}+\frac{1}{2}-\frac{1}{2\nu})^\nu)^{1/\nu} +\frac{1}{2\nu}-\frac{1}{2} \nonumber \\
 &=& e_{xt}\mu_{xt}+\frac{1}{2}-\frac{1}{2\nu}+\frac{1}{2\nu}-\frac{1}{2} \nonumber \\
 &=& e_{xt}\mu_{xt},
\label{eqn:cmp_dxt_mean_var}
\end{eqnarray}
as implied from (\ref{eqn:cmp_mean_var}).

We remark that it may be sufficient to specify $D_{xt} \vert \mu_{xt},\nu \sim \text{CMP }\left((e_{xt}\mu_{xt})^\nu , \nu\right)$ in order to achieve $\mathbb{E}[D_{xt} ] \approx e_{xt}\mu_{xt}$, particularly since $\nu$ is typically not too small. Nevertheless, we proceed with the formulation in (\ref{eqn:cmp_dxt}) as it offers a more accurate adjustment without introducing significant additional complexity.

The specification also implies that the variance of the number of deaths, expressed in terms of mean, is given by
\begin{eqnarray}
\text{Var}[D_{xt} ] &\approx& \frac{((e_{xt}\mu_{xt}+\frac{1}{2}-\frac{1}{2\nu})^\nu)^{1/\nu}}{\nu} \nonumber \\
 &=& \frac{e_{xt}\mu_{xt}+\frac{1}{2}-\frac{1}{2\nu}}{\nu} \nonumber \\
 &=& \frac{\mathbb{E}[D_{xt}]+\frac{1}{2}-\frac{1}{2\nu}}{\nu}.
 \label{eqn:var_dxt_cmp}
\end{eqnarray}
So $\nu$ governs the level of overdispersion, where the term $1/\nu$ can also be (approximately) interpreted as the multiplicative increase in the variance of $D_{xt}$ relative to its mean. 
For instance, if $\nu < 1$, we have $\text{Var}[D_{xt} \vert \mu_{xt},\nu] > \mathbb{E}[D_{xt}\vert \mu_{xt},\nu]$ which is the case for overdispersion, and vice versa. 
And when $\nu = 1$, $\text{Var}[D_{xt}\vert \mu_{xt},\nu] = \mathbb{E}[D_{xt}\vert \mu_{xt},\nu]$ indicating the case of equidispersion. 

Similar to before, we consider two structural models for $\mu_{xt}$,
\begin{equation}
\log \mu_{xt} =
\begin{cases}
\alpha_x + \beta_x \kappa_t, \\
\alpha_x + \beta_x \kappa_t + \gamma_c,
\end{cases}
\label{eqn:rate_model}
\end{equation}
coupled with the use of (\ref{eqn:cmp_dxt}). We term the former the \textbf{CMP--LC} model, and the latter as the \textbf{CMP--LCC} model.

Using the property of independence and the PMF above, we also observe that the full data likelihood is
\begin{equation}
f(\boldsymbol{d}|\boldsymbol{\alpha},\boldsymbol{\beta},\boldsymbol{\kappa},\nu)=\prod_{x,t}\mathbb{P}[D_{xt}=d_{xt}|\mu_{xt},\nu] =\prod_{x,t} \frac{1}{Z(\lambda_{xt},\nu)}\frac{(\lambda_{xt})^{d_{xt}}}{(d_{xt}!)^\nu}.
\label{eqn:cmp_likelihood}
\end{equation}


\section{Bayesian estimation}
\label{sec:bayesian}

The motivation for considering a Bayesian framework has been extensively discussed in the literature (e.g., \citealp{wong_2017}) . In our setting, this approach allows the dispersion parameter $\nu$ to be treated as an unknown random variable to be estimated naturally and directly from the data. This data-driven specification enables inference on the level of dispersion without the need to prespecify a fixed value of $\nu$. This also forms a coherent framework for uncertainty calibration, including that due to overdispersion and other sources of uncertainty, such as parameter uncertainty through prior distributions. In the following subsections, we describe the prior specification, derive the joint posterior distributions, and outline our proposed MCMC sampling procedure. 

\subsection{Prior specification}
\label{sec:prior}

In principle, the specified prior distributions should reflect our prior knowledge about the model parameters. But in most applications, we do not have prior domain knowledge, so Bayesian methods allow us to learn about all parameters (including $\nu$) from the data. In this paper, we will predominantly follow the specification suggested by \cite{wong_2023}, except for $\nu$ where we explore various specifications. 

First, we specify
\begin{align*}
\vect{\alpha}&\sim \hspace{0.3cm} N(-5\cdot\vect{1}_{A},4\cdot{I}_{A}), \\
\vect{\beta}&\sim \hspace{0.3cm} N\left(\frac{1}{A}\cdot\vect{1}_{A-1},0.005\cdot\left({I}_{A-1}-\frac{1}{A}\cdot{J}_{A-1}\right)\right), \\
\beta_1 &= 1-\sum_{x=2}^{A}\beta_x, 
\end{align*}
where $\vect{\alpha}=(\alpha_1,\ldots,\alpha_A)^\top$, $\vect{\beta}=(\beta_2,\ldots,\beta_A)^\top$, $\vect{1}_{A}$ is a length$-A$ vector of ones, ${I}_{A}$ and ${J}_{A}$ are respectively the identity matrix and a matrix of ones with dimension $A\times A$.

Following the formulation of \cite{bayespoisson}, we impose for $\kappa_t$
\begin{eqnarray}
\left.\begin{array}{l c}
\kappa_{t}-\eta_{t}=\rho(\kappa_{t-1}-\eta_{t-1})+\epsilon_{t} , & \mbox{ for } t=2,3,\ldots,T, \\
\kappa_1=\eta_1+\epsilon_1, & \end{array}\right\}
\label{eqn:ar1}
\end{eqnarray}
where $\eta_{t}=\psi_1+\psi_2 t$ denotes the linear drift and $\epsilon_t\stackrel{{\rm ind}}{\sim} N(0,\sigma_{\kappa}^2)$. Then, depending on the value of $\rho$, this forms either a first-order autoregressive (AR(1)) model ($|\rho|<1$) or a random walk model ($\rho=1$) for $\kappa_t$. The model in (\ref{eqn:ar1}) can be expressed multivariately (along with the constraint) as
\begin{eqnarray}
\left.\begin{array}{l}\boldsymbol{\kappa}|\rho,\vect{\psi},\sigma_\kappa^2 \sim N\left(\vect{\mu}_\kappa,\sigma_\kappa^2 {V}\right), \\
\kappa_1=-\sum_{t=2}^{T} \kappa_t, \end{array}\right\}
\label{eqn:mult_ar1}
\end{eqnarray}
where $\vect{\kappa}=(\kappa_2,\ldots,\kappa_T)^\top$, $\vect{\mu}_\kappa=({I}_{T-1}-{B}_{21}{B}_{11}^{-1}\vect{1}_{T-1}^\top)\cdot{X}\boldsymbol{\psi}$,
${X}=\left(\begin{array}{c c c c}
1 & 1 & \cdots & 1 \\
2 & 3 & \cdots & T  \end{array}\right)^\top\, ,$ $\vect{\psi}=(\psi_1,\psi_2)^\top$, expressions for ${B}$ and ${V}$ are provided in Appendix A. For other hyperparameters,
\begin{align*}
\frac{\rho+1}{2} &\sim \hspace{0.3cm} {\rm Beta}(3,2), \mbox{ where $\rho \in (-1,1)$}, \\
\sigma_\kappa^{-2} &\sim \hspace{0.3cm} \mbox{Gamma}(1,0.0001), \\
\vect{\psi} &\sim \hspace{0.3cm} N\left(\left(\begin{array}{c}
0 \\
0
\end{array}\right),\left(\begin{array}{c c}
2000 & 0 \\
0 & 2
\end{array}\right)\right), 
\end{align*}

For the cohort parameters $\gamma_c$, we use an ARIMA(1,1,0) as adopted by \cite{stmomo}, i.e.
\begin{eqnarray}
\left.\begin{array}{l l}
(\gamma_c-\gamma_{c-1})=\rho_\gamma (\gamma_{c-1}-\gamma_{c-2})+\epsilon_c^\gamma , & \mbox{ for } c=3,\ldots,C, \\
\gamma_2-\gamma_1=\frac{1}{\sqrt{1-\rho_\gamma^2}}\epsilon_2^\gamma, & \\
\gamma_1= 100\epsilon_1^\gamma,
\end{array}\right\}
\label{eqn:gamma_projection_model}
\end{eqnarray}
where $\epsilon_c^{\gamma}\stackrel{{\rm ind}}{\sim} N(0,\sigma_\gamma^2)$ for $c=1,\ldots,C$, and $\rho_\gamma$ and $\sigma_\gamma^2$ are hyperparameters. Applying the constraints $\sum_c \gamma_c=\sum_c c\gamma_c=\sum_c c^2\gamma_c=0$ on (\ref{eqn:gamma_projection_model}), we write
\begin{eqnarray}
\vect{\gamma}\sim N(\vect{0},\sigma_\gamma^2{W}_\gamma),
\label{eqn:gamma_prior_withcons}
\end{eqnarray}
where $\vect{\gamma}=(\gamma_2,\ldots,\gamma_{71},\gamma_{73},\ldots,\gamma_{C-1})^\top$, and ${W}_\gamma$ is provided in Appendix A. The remaining cohort components $\{\gamma_1,\gamma_{72},\gamma_C\}$ can then be derived as follows
\begin{eqnarray*}
\gamma_1 &=& \frac{1}{71\times (C-1)}\sum_{c\neq 1,72,C} (c-72)(C-c)\gamma_c, \\
\gamma_{72} &=& -\frac{1}{69\times 71}\sum_{c\neq 1,72,C} (C-c)(c-1)\gamma_c, \\
\gamma_C &=& \frac{1}{69\times (C-1)}\sum_{c\neq 1,72,C} (c-1)(72-c)\gamma_c.
\end{eqnarray*}
As recommended by \cite{wong_2023}, we set priors
$$\rho_\gamma\sim N(0,1) \mbox{\hspace{1cm} and \hspace{1cm}} \sigma_\gamma\sim \mbox{Uniform}(0.1).$$

For the dispersion parameter $\nu$, our initial strategy is to specify (improper) flat prior, i.e. 
\begin{equation}
\pi(\nu)\propto 1 \ , \text{ for }\nu>0
\label{eqn:prior_nu}
\end{equation}
so that results derived from our Bayesian analysis are directly comparable with those from a frequentist framework. 
This is in the spirit of data-driven inferences, where we allow the data to ``speak for itself''.
This also serves as an indication if there is sufficient information in the data to estimate $\nu$.
We remark that the use of improper flat priors is common and convenient, which typically leads to proper posteriors, as pointed out by \cite{gelfand1999identifiability}.

It is also of interest to determine the influence of prior specification on the results. 
We could explore different choices of $\pi(\nu)$. 
We focus on priors of the form $$\pi(\nu)\sim \text{Gamma}(a,b),$$ 
where $a$ and $b$ are hyperparameters to be specified as follows:

\begin{itemize}
  \item $a=1$ and $b=100$.
   \item $a=1$ and $b=10$.
 \item $a=1$ and $b=1$.
 \item $a=1$ and $b=0.1$.
 \item $a=1$ and $b=0.01$.
\end{itemize}

From top to bottom, the prior variance of $\nu$ increases from $1\times 10^{-4}$ to $1\times 10^{4}$, indicating that the priors become increasingly vague or non-informative. A sensitivity analysis with respect to these values of $a$ and $b$ will be presented later in Section \ref{sec:results}.

\subsection{Posterior distributions}

First, denote $\boldsymbol{\theta}=\{\boldsymbol{\alpha},\boldsymbol{\beta},\boldsymbol{\kappa},\vect{\psi},\sigma_\kappa^2,\rho,\nu\}$ as the full set of parameter for the CMP--LC model, the joint posterior distribution is then
\begin{eqnarray*}
\pi(\boldsymbol{\theta}|\boldsymbol{d}) &\propto & f(\boldsymbol{d}|\boldsymbol{\alpha},\boldsymbol{\beta},\boldsymbol{\kappa},\nu) \cdot \pi(\boldsymbol{\theta}) \\ 
&=& f(\boldsymbol{d}|\boldsymbol{\alpha},\boldsymbol{\beta},\boldsymbol{\kappa},\nu) \cdot \pi(\boldsymbol{\alpha}) \cdot \pi(\boldsymbol{\beta}) \cdot \pi(\boldsymbol{\kappa}|\vect{\psi},\sigma_\kappa^2,\rho) \cdot \pi(\vect{\psi}) \cdot \pi(\sigma_\kappa^2)\cdot \pi(\rho) \cdot \pi(\nu),
\end{eqnarray*}
where $f(\boldsymbol{d}|\boldsymbol{\alpha},\boldsymbol{\beta},\boldsymbol{\kappa},\nu)$ is the data likelihood as given in (\ref{eqn:cmp_likelihood}), and the rest are prior distributions of all other parameters as described in Section \ref{sec:prior}. 
Similarly, for the CMP--LCC model, $\boldsymbol{\theta}=\{\boldsymbol{\alpha},\boldsymbol{\beta},\boldsymbol{\kappa},\vect{\psi},\sigma_\kappa^2,\rho,\boldsymbol{\gamma},\sigma_\gamma^2,\rho_\gamma,\nu\}$, giving the joint posterior
\begin{eqnarray*}
\pi(\boldsymbol{\theta}|\boldsymbol{d}) &\propto& f(\boldsymbol{d}|\boldsymbol{\alpha},\boldsymbol{\beta},\boldsymbol{\kappa},\nu) \cdot \pi(\boldsymbol{\alpha}) \cdot \pi(\boldsymbol{\beta}) \cdot \pi(\boldsymbol{\kappa}|\vect{\psi},\sigma_\kappa^2,\rho) \cdot \pi(\vect{\psi}) \cdot \pi(\sigma_\kappa^2) \\
&& \cdot \pi(\rho)\cdot \pi(\boldsymbol{\gamma}|\sigma_\gamma^2,\rho_\gamma)\cdot \pi(\sigma_\gamma^2) \cdot \pi(\rho_\gamma) \cdot \pi(\nu).
\end{eqnarray*}

\subsection{The MCMC updating scheme}

We apply the variable-at-a-time random walk Metropolis-Hastings (MH) algorithm as described in \cite{kendall}. For a description of the MCMC updating schemes of all other parameters please refer to Appendix B in the Supplementary Materials. Here we describe in detail the MCMC updating step for the dispersion parameter, $\nu$. 

Note that the conditional posterior distribution of $\nu$ is proportional to the likelihood function multiplied by their prior distributions, i.e. 
\begin{equation}
\pi(\nu \vert \boldsymbol{d},\boldsymbol{\theta}_{-\nu}) \propto f(\boldsymbol{d}|\boldsymbol{\alpha},\boldsymbol{\beta},\boldsymbol{\kappa},\nu)  \times \pi(\nu), 
\label{eqn:cond_post_nu}
\end{equation}
where $\boldsymbol{\theta}_{-\nu}=\boldsymbol{\theta}\backslash\{\nu\}=\{\boldsymbol{\alpha},\boldsymbol{\beta},\boldsymbol{\kappa},\vect{\psi},\sigma_\kappa^2,\rho\}$ is the full set of parameters excluding $\nu$. MCMC updating scheme is then applied iteratively to sample from (\ref{eqn:cond_post_nu}), which can be combined with other posterior samples to form samples of the joint posterior distribution, $\pi(\boldsymbol{\theta} \vert \boldsymbol{d})$.

To avoid proposing negative values, we perform the updating scheme on a log scale, i. e. 
$$\log (\nu^*) \sim N(\log (\nu^{(i)}), \sigma^2_\nu),$$
where $\nu^{(i)}$ is the current iterate, $\sigma_\nu^2$ is the proposal variance chosen so that the resulting acceptance rate is between 0.15-0.45 (\citealp{acceptrate}). From our pilot runs, setting $\sigma_\nu^2=0.005$ leads to an acceptance rate of about 0.40.

The acceptance ratio for $\nu^*$ is
\begin{equation}
\alpha(\nu^* \vert \boldsymbol{\theta}_{-\nu}^{(i)},\nu^{(i)})=\min\left\{1,\frac{f(\boldsymbol{d}|\boldsymbol{\alpha}^{(i)},\boldsymbol{\beta}^{(i)},\boldsymbol{\kappa}^{(i)},\nu^*)  \times \pi(\nu^*)}{f(\boldsymbol{d}|\boldsymbol{\alpha}^{(i)},\boldsymbol{\beta}^{(i)},\boldsymbol{\kappa}^{(i)},\nu^{(i)})  \times \pi(\nu^{(i)})}\right\}. 
\label{eqn:accept_nu}
\end{equation}
If the flat prior in (\ref{eqn:prior_nu}) is used, then the expression in (\ref{eqn:accept_nu}) reduces to simply a ratio of likelihoods.

\subsection{Initialisation and Convergence Diagnostics}
\label{sec:initialization}
    For the initialisation of $\vect{\alpha}$, $\vect{\beta}$, $\vect{\kappa}$ and $\vect{\gamma}$, we use the MLE obtained by iteratively fitting the Poisson--LCC model using the \textit{glm} function in R. The initial values of $\sigma_{\kappa}^{2}$ and $\rho$ are obtained by fitting an AR(1) with linear drift model on $\vect{\kappa}$ (using the \textit{arima} function within the \textit{forecast} package in \R). The cohort-related parameters, $\sigma_\gamma$ and $\rho_\gamma$, are initialised by fitting an ARIMA (1,1,0) model to the MLE of $\vect{\gamma}$. Finally, $\vect{\psi}$ is initialised as $(0,0)^{\top}$, while the dispersion parameter is initialised as $\nu=0.5$. Note that the initialization is proposed on the basis of values close to the MLE to speed up convergence, but it should have little effect on the estimation procedure. 

We apply a burn-in phase of $1{,}000$ iterations to mitigate the effect of initialization. In addition, we apply 50$^{\rm th}$ posterior sample thinning (collecting one realization every 50 iterations) to each of the parameters to reduce the autocorrelations of these series. After discarding burn-in iterations and applying thinning, we obtain a sample of size $10{,}000$ for each of the parameters under both the CMP--LC and CMP--LCC models.  	

Before making any inferential comparisons, trace plots and auto-correlation plots (see for example \citealp[Chapter~3]{bugs}) can be used as diagnostic tools for detecting anomalies in the MCMC generated posterior samples. By referring to Appendix C, the trace plots of some of the randomly selected parameters emerge as if convergence has been attained, with proper mixing and no apparent anomaly. The sample auto-correlations also appear to decay fairly quickly after applying thinning, except perhaps $\kappa_t$, which are relatively more correlated. A few commonly used convergence diagnostics such as those proposed by \cite{rubin} and \cite{geweke1991evaluating}, all of which are presented in the Supplementary Materials. Specifically, the Gelman's statistic, potential scale reduction factor (PSRF) computed from the posterior samples under both the CMP--LC and CMP--LCC are all very close to 1, indicating that there is no convergence issue. The Z statistics computed using the approach by \cite{geweke1991evaluating} also satisfy the convergence criterion.

In summary, the MCMC generated posterior samples seem to be well-behaved and, thus, are ready to be used to perform subsequent computations for accurate inferences to be drawn.

\section{Results}
\label{sec:results}

When presenting our results, we also include, where appropriate, those obtained under the Poisson and negative binomial specifications for $D_{xt}$ using Bayesian methods, for comparison purposes (see \citealp{bayespoisson} and \citealp{wong_2017}, respectively). The Poisson specification is as described in Section \ref{sec:plc} and the negative binomial (NB) as follows, 
\begin{equation}
    D_{xt}\sim \text{Neg-Bin}\left(\phi,\frac{\phi}{e_{xt}\mu_{xt}+\phi}\right),
\end{equation}
where $\phi$ is the overdispersion parameter. Under the NB specification, $\mathbb{E}[D_{xt}]=e_{xt}\mu_{xt}$ and $\text{Var}[D_{xt}]=\mathbb{E}[D_{xt}]\times (1+\mathbb{E}[D_{xt}]/\phi)>\mathbb{E}[D_{xt}]$. The comparison is intended to highlight firstly the significance of accounting for overdispersion, and secondly the difference between two variants of incorporating overdispersion (i.\,e. between the CMP and NB distributions).

\subsection{Estimated parameters}

\begin{figure}[!htbp]
\centering\includegraphics[scale=0.9]{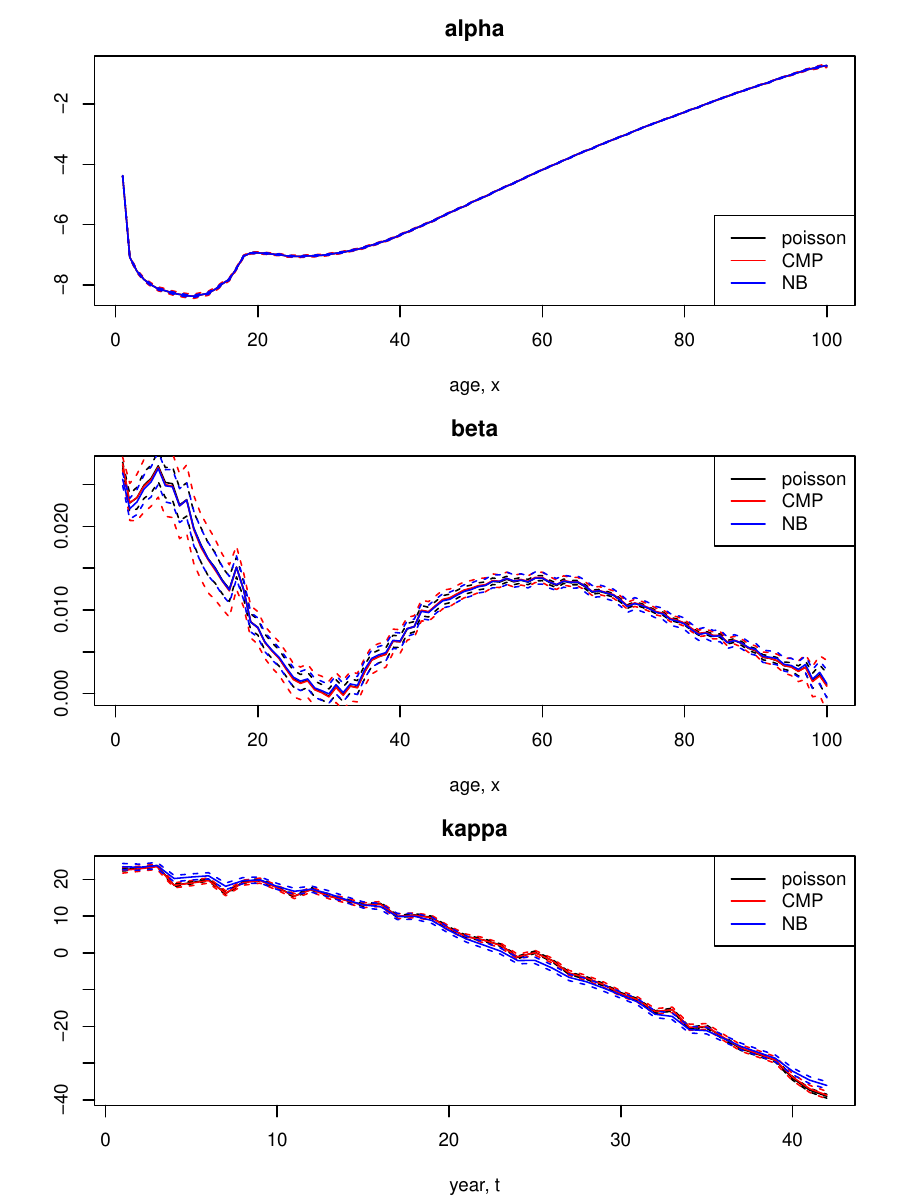}
\caption{\label{fig:posterior_abk_LC}Plot of the medians (solid lines) and $95\%$ credible intervals (dashed lines) of the estimated model parameters under the Poisson--LC, CMP--LC, and NB--LC models.}
\end{figure}

\begin{figure}[!htbp]
\centering\includegraphics[scale=0.9]{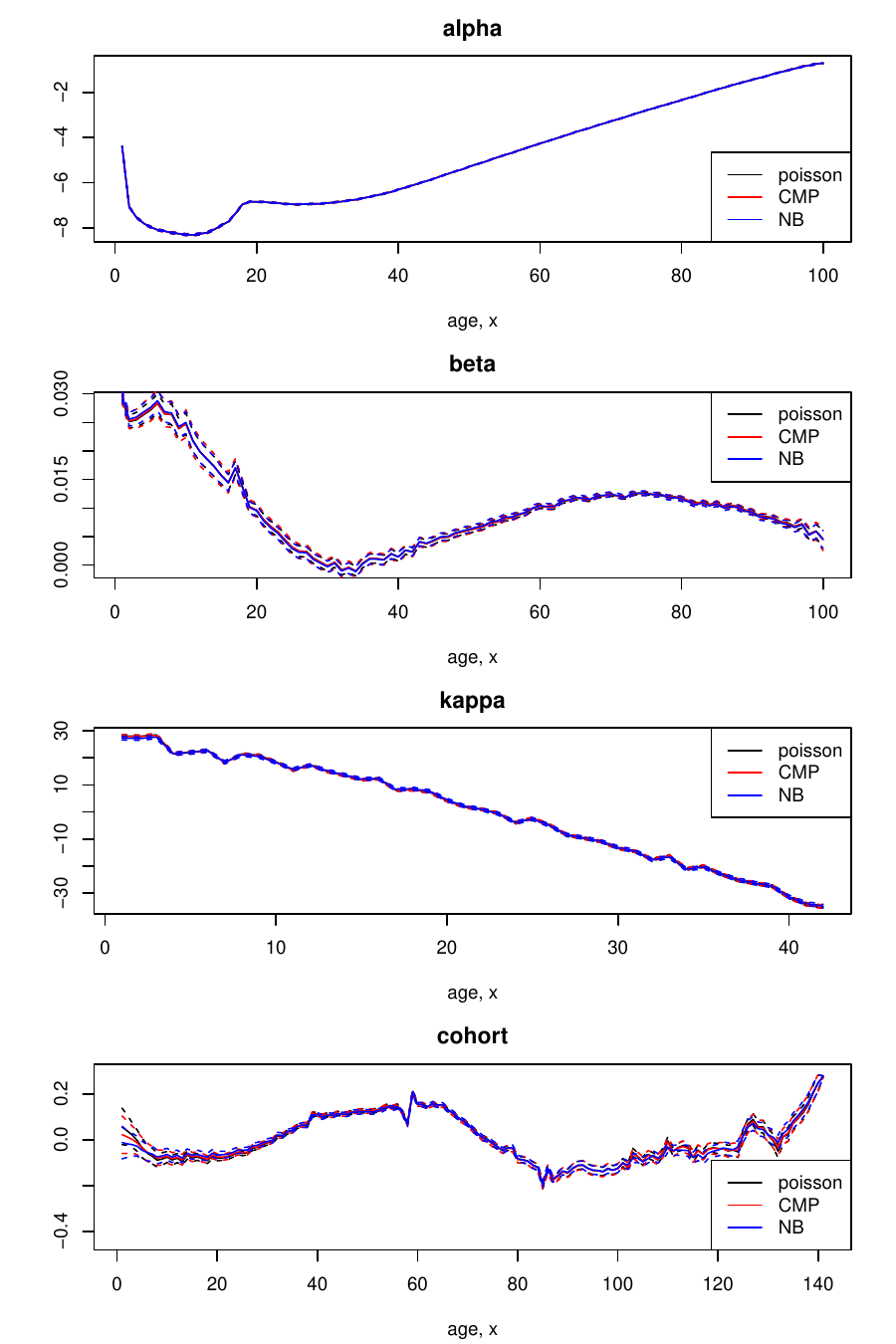}
\caption{\label{fig:posterior_abk_LCC}Plot of the medians (solid lines) and $95\%$ credible intervals (dashed lines) of the estimated model parameters under the Poisson--LCC, CMP--LCC, and NB--LCC models.}
\end{figure}

Figures \ref{fig:posterior_abk_LC} and \ref{fig:posterior_abk_LCC} depict the fitted values (posterior medians) of $\vect{\alpha}$, $\vect{\beta}$ and $\vect{\kappa}$, accompanied by the associated $95\%$ credible intervals under respectively the LC and LCC models (for CMP, Poisson, and NB specifications). Evidently, the fitted values under these models are rather similar, with the overdispersion models producing slightly wider credible intervals. This is the general feature of models which account for overdispersion, where the responses ($D_{xt}$) are allowed to have more variabilities due to the extra flexibility offered by the model likelihood, permitting the parameters to be more volatile, and hence, the wider credible intervals. Some exceptions observed for $\vect{\beta}$ in the CMP--LC model, where the CMP specification yields credible intervals that are noticeably wider. $\vect{\kappa}$ for the NB--LC model also appear to demonstrate more linearity than the CMP--LC model.

\begin{figure}[!htbp]
\centering\includegraphics[scale=0.9]{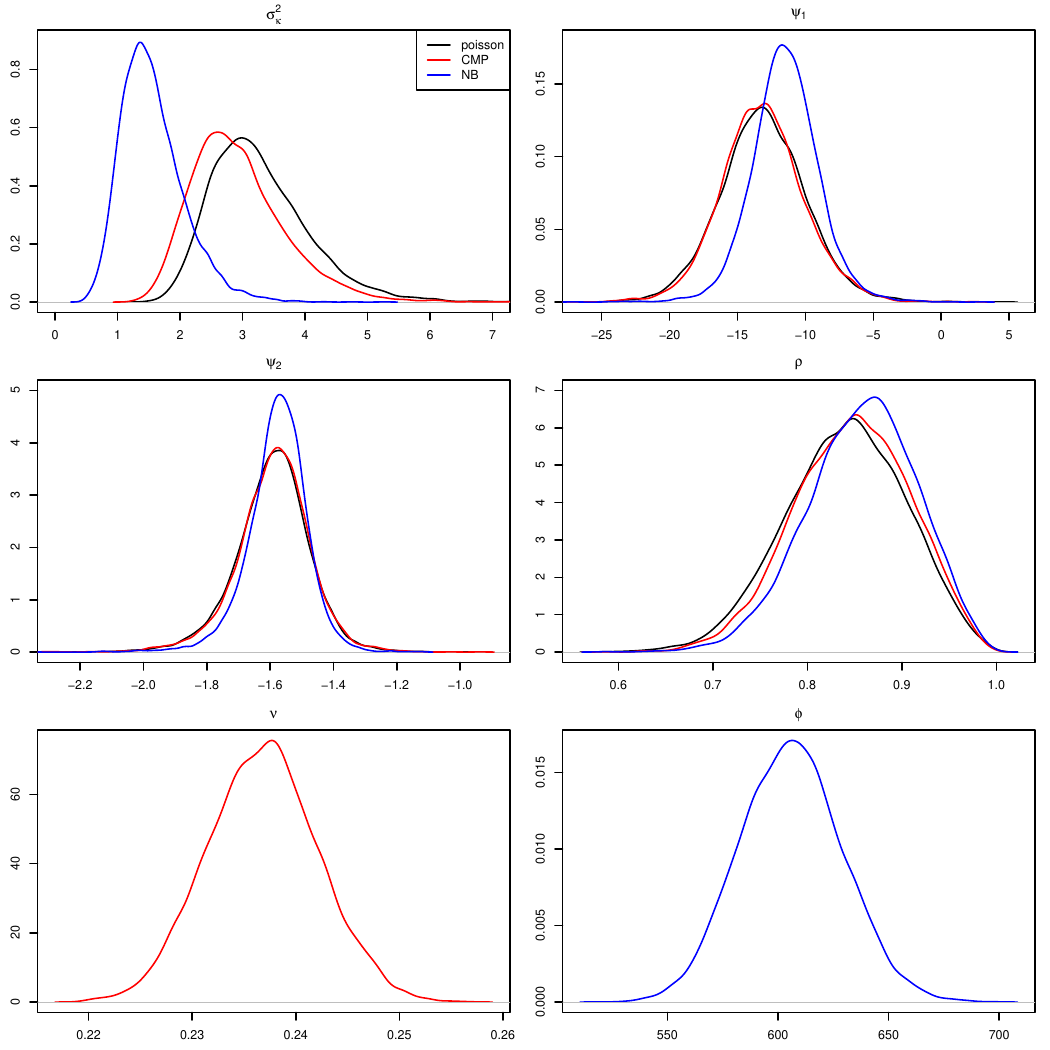}
\caption{\label{fig:density_LC}Density plots of estimated parameters for the LC models.}
\end{figure}

\begin{figure}[!htbp]
\centering\includegraphics[scale=0.9]{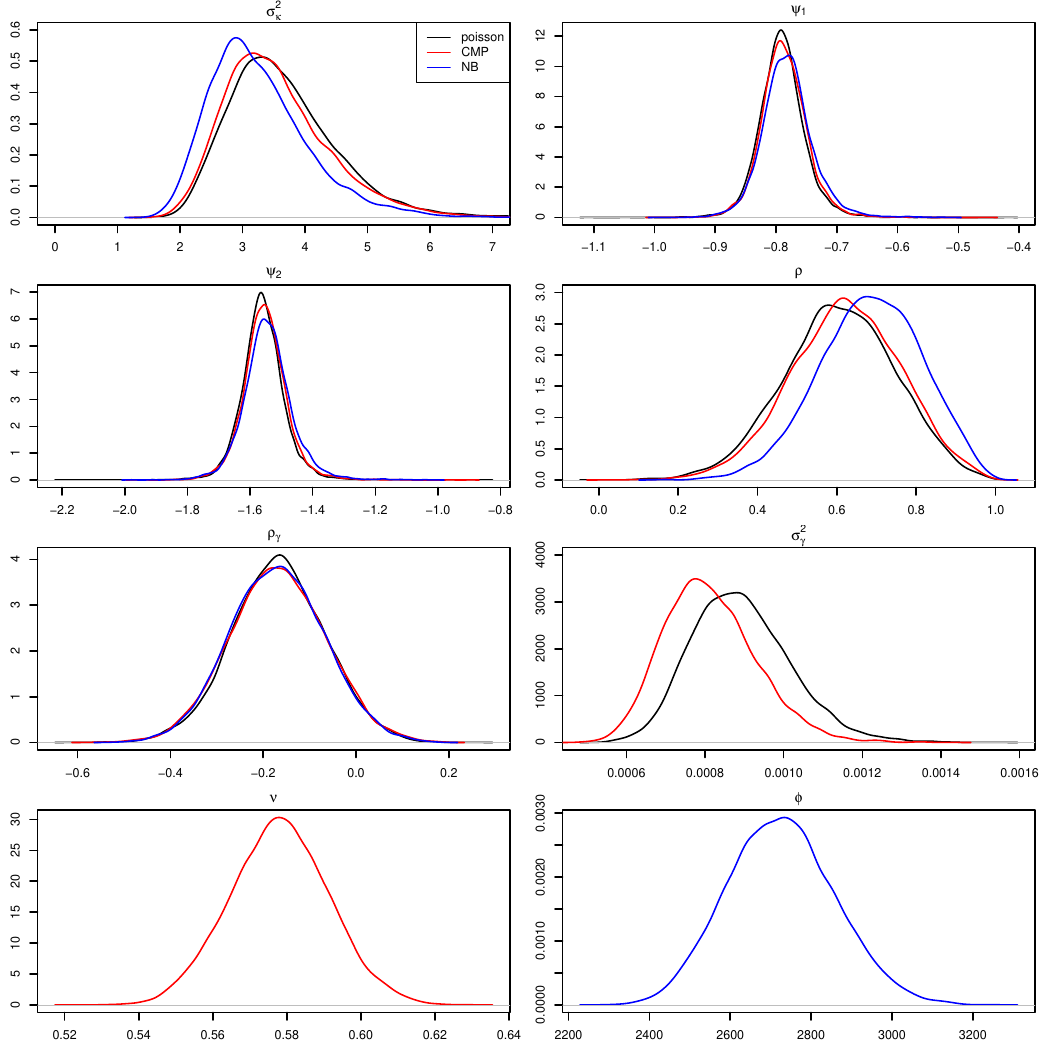}
\caption{\label{fig:density_LCC}Density plots of estimated parameters for the LCC models.}
\end{figure}

The kernel posterior densities of other parameters for the LC and LCC models are illustrated in Figures \ref{fig:density_LC} and \ref{fig:density_LCC}. The posteriors are mostly similar between the CMP and Poisson specifications. Between the CMP and NB specifications, they are rather similar (visually) except for $\sigma_\kappa^2$, $\psi_1$, and $\rho$. In particular, the estimated $\sigma_\kappa^2$ are smaller, while the estimated $\psi_1$ and $\rho$ are larger for the NB specification. Interestingly, the posteriors of $\rho$ are uni-modal and have values close to one, contrary to the findings of \cite{wong_2017} for England and Wales female death data where $\rho$ demonstrated bimodal posteriors. 

The estimated mean dispersion parameter $\nu$ under the CMP--LC and CMP--LCC models are respectively around $0.237$ and $0.578$. 
Recall from (\ref{eqn:var_dxt_cmp}) that $1/\nu$ measures the relative (multiplicative) increase in the variance of $D_{xt}$ with respect to its mean. Substituting the posterior means into the expression gives $\frac{1}{\nu}\approx 4.22$ \footnote[2]{This means that $\text{Var}[D_{xt}]$ is approximately 4.22 times $\mathbb{E}[D_{xt}]$.} and  $\frac{1}{\nu}\approx 1.73$, indicating that the level overdispersion implied by the CMP--LCC is lower than that by the CMP--LC model. This finding is consistent with that by \cite{wong_2023}, where incorporating cohort effects allows for better calibration between data signals and uncertainties, helping to prevent misclassification of uncaptured data signals as overdispersion.


\subsection{Model assessments}

\begin{figure}[htbp]
\includegraphics[scale=0.95]{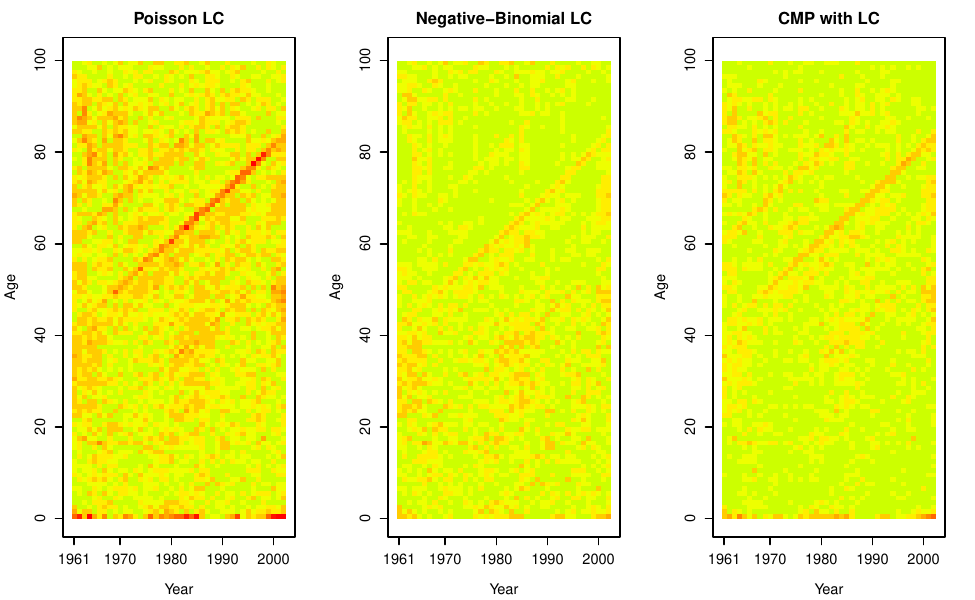}
\caption{Heatmap of $r_{xt}^2$ for the LC models.}
\label{fig:heatmap_compare_LC_male}
\end{figure}

\begin{figure}[htbp]
\includegraphics[scale=0.95]{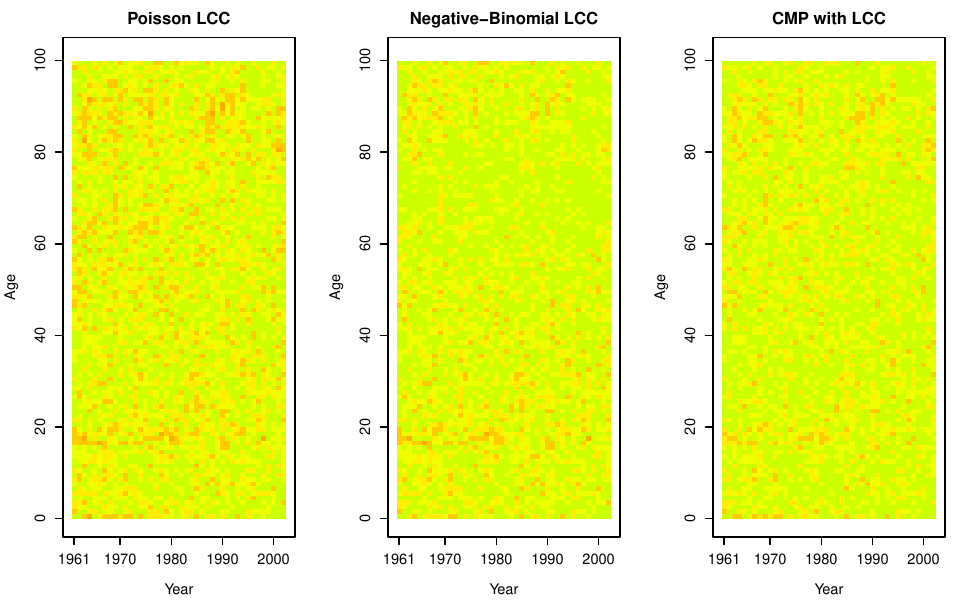}
\caption{Heatmap of $r_{xt}^2$ for the LCC models.}
\label{fig:heatmap_compare_LCC_male}
\end{figure}

Heatmaps of residuals by age and time are presented in Figures \ref{fig:heatmap_compare_LC_male} and \ref{fig:heatmap_compare_LCC_male} to demonstrate in-sample goodness of fit. In general, both the CMP and NB specifications yield residuals that are well-behaved, indicating that the error distributions are well-calibrated, hence, model adequacy. By contrast, the Poisson specification clearly led to poor distribution of residuals, highlighting its inadequacy. Comparing the CMP and NB specifications, the NB appears to capture variations more effectively in ages 60–80 but struggles at younger ages, whereas the CMP performs better at younger ages. Overall, the residual patterns suggest that the CMP specification provides superior fit.

Table \ref{tab:pearson_residuals} reports the total model residuals $r^2$, as defined in Section \ref{sec:overdispersion}, for assessing overall goodness of fit. Evidently, the CMP specification outperforms its counterparts by producing smaller overall deviance.

\begin{table}[!htbp]
\caption{The estimated total model residuals $r^2$ under each of the model specifications. Also included in parentheses are the percentages of cells with poor fit, i.e. $r^2_{xt}>3.84$. Models that generate less than $5\%$ poor fit cells are good.}
  \label{tab:pearson_residuals}
  \centering
\begin{tabular}{r rrr}
\hline
Model & Poisson & NB & CMP \\
\hline
LC & $16709.85$ & $4392.80$ & $\vect{3970.60}$ \\
 & $(26.86\%)$ & $(5.62\%)$ & $\boldsymbol{(4.40\%)}$ \\
 \hline
LCC & $6628.97$ & $4006.45$ & $\vect{3850.75}$ \\
 & $(11.29\%)$ & $(4.71\%)$ & $\boldsymbol{(4.33\%)}$ \\
\hline
\end{tabular}
\end{table}

The out-of-sample predictive performances of the models are compared by assessing their logarithmic scores (LOGS), continuous ranked probability scores (CRPS), and Dawid-Sebastiani scores (DSS), see for example \cite{MCMC_scoring_2021}. These measures can be derived from the MCMC posterior samples of $D_{xt}$ (for $x=1,\ldots,A$ and $t=2003,\ldots,2021$) and computed using the R package \textit{scoringRules} (\citealp{scoringrules_package}), as presented in Table \ref{tab:scores}. The results indicated that the CMP specifications for death counts generated better out-of-sample forecasts than the counterparts.

\begin{table}[htbp]
\caption{Various scores for assessing out-of-sample predictive performances under each of the model specifications (the smaller the better). }
  \label{tab:scores}
  \centering
\begin{tabular}{r r rrr}
\hline
Model & Metric & Poisson & NB & CMP \\
\hline
LC & LOGS & $10.82$ & $9.05$ & $\vect{7.20}$ \\
 & CRPS & $235.62$ & $254.97$ & $\vect{227.60}$ \\
 & DSS & $15.37$ & $13.91$ & $\vect{12.71}$ \\
 \hline
LCC & LOGS & $15.57$ & $13.86$ & $\vect{8.95}$ \\
 & CRPS & $153.95$ & $153.32$ & $\vect{151.85}$ \\
 & DSS & $15.34$ & $14.32$ & $\vect{13.29}$ \\
\hline
\end{tabular}
\end{table}

\subsection{Fitted and projected crude mortality rates}
\label{sec:fitted_projected_rates}

We also assess the performance of the models using crude mortality rates (observable) rather than the underlying mortality rates, $\mu_{xt}$ (unobservable). To obtain the fitted crude mortality rates, the fitted number of deaths, $D_{xt}^F$, is first generated through Equation (\ref{eqn:cmp_dxt}), where joint posterior samples of $\alpha_x$, $\beta_x$, $\kappa_t$, $\gamma_c$, and $\nu$ are substituted as appropriate. The crude fitted mortality rate for each $x=1,\ldots,A$ and $t=1,\ldots,T$ is then computed as $\hat{\mu}_{xt}=\frac{D_{xt}^F}{e_{xt}}$.

For mortality projection, we note that the posterior predictive distribution of 1-year ahead number of deaths for each age (with age parameters held fixed), under the LCC model for instance, is
\begin{eqnarray}
&& f(D_{x\,T+1}|\vect{d}) \nonumber \\ 
&=&\int f(D_{x\, T+1}|\alpha_x,\beta_x,\kappa_{T+1},\gamma_{T+1-x},\nu)\times f(\kappa_{T+1}|\kappa_T,\vect{\psi},\rho,\sigma_\kappa^2)\times f(\gamma_{T+1-x}|\gamma_{T-x},\gamma_{T-x-1},\rho_\gamma,\sigma_\gamma^2) \nonumber\\
&&\times \pi(\alpha_x,\beta_x,\kappa_T,\vect{\psi},\rho,\sigma_\kappa^2,\gamma_{T+1-x},\rho_\gamma,\sigma_\gamma^2,\nu|\vect{d}) {\rm d}\alpha_x \ldots {\rm d}\phi,
\end{eqnarray}
where $\pi(\alpha_x,\beta_x,\kappa_T,\vect{\psi},\rho,\sigma_\kappa^2,\gamma_{T+1-x},\rho_\gamma,\sigma_\gamma^2,\nu|\vect{d})$ is the joint posterior distribution. This suggests the following generation procedure for the projected crude mortality rates:

\begin{enumerate}[(i)]
\item Generate $\kappa_{T+1}$ from Equation (\ref{eqn:ar1}) where joint posterior samples of $(\vect{\psi},\rho,\kappa_T,\sigma_\kappa^2)$ are appropriately substituted.

\item Generate $\gamma_{T}$ from Equation (\ref{eqn:gamma_projection_model}) where joint posterior samples of $(\gamma_{T-1},\gamma_{T-2},\rho_\gamma,\sigma_\gamma^2)$ are substituted. Other ``observed'' cohort components $(\gamma_{T-99},\ldots,\gamma_{T-3})$ are also available from posterior samples.

\item Generate for each $x$ the forecasted number of deaths using (\ref{eqn:cmp_dxt}), i.e.,
$$D^F_{x\,T+1} \sim \text{CMP }\left((e_{x\, T+1}\exp(\alpha_x+\beta_x\kappa_{T+1}+\gamma_{T+1-x})+\frac{1}{2}-\frac{1}{2\nu})^\nu , \nu\right),$$
where $e_{x\, T+1}$ are holdout central exposed to risks, joint posterior samples of $(\alpha_x,\beta_x,\nu)$ are substituted, $\kappa_{T+1}$ and $\gamma_{T+1-x}$ are respectively from steps (i) and (ii).

\item Compute for each $x$ the projected crude mortality rates as $\hat{\mu}_{x\, T+1}=\frac{D^F_{x\,T+1}}{e_{x\,T+1}}$.
\end{enumerate}
By analogy, $h$-year ahead projections can be obtained by recursive implementation of the above algorithm. Having generated samples for $\hat{\mu}_{xt}$ for $x=1,\ldots,A$ and $t=1,\ldots,T+h$, sample percentiles can be constructed to form median and 95$\%$ intervals. 

Under the Poisson and NB specifications, we generate from $D_{xt}^F\sim \mbox{ Poisson}(e_{xt}\mu_{xt})$ and $D_{xt}^F\sim \mbox{ Neg-Bin}(\phi,\frac{\phi}{e_{xt}\mu_{xt}+\phi})$, using posterior samples of the relevant parameters, and then compute the fitted crude mortality rates as $\hat{\mu}_{xt}=\frac{D_{xt}^F}{e_{xt}}$.  





       
Note that the above generation procedure of fitted and projected mortality rates builds in uncertainties due to random variations (Poisson, CMP, and NB), parameter, and forecast uncertainties. This allows us to better visualise the overall performances of each distributional specification for $D_{xt}$, taking into account various sources of uncertainties.

\begin{figure}[htbp]
\centering
\includegraphics[scale=0.9]{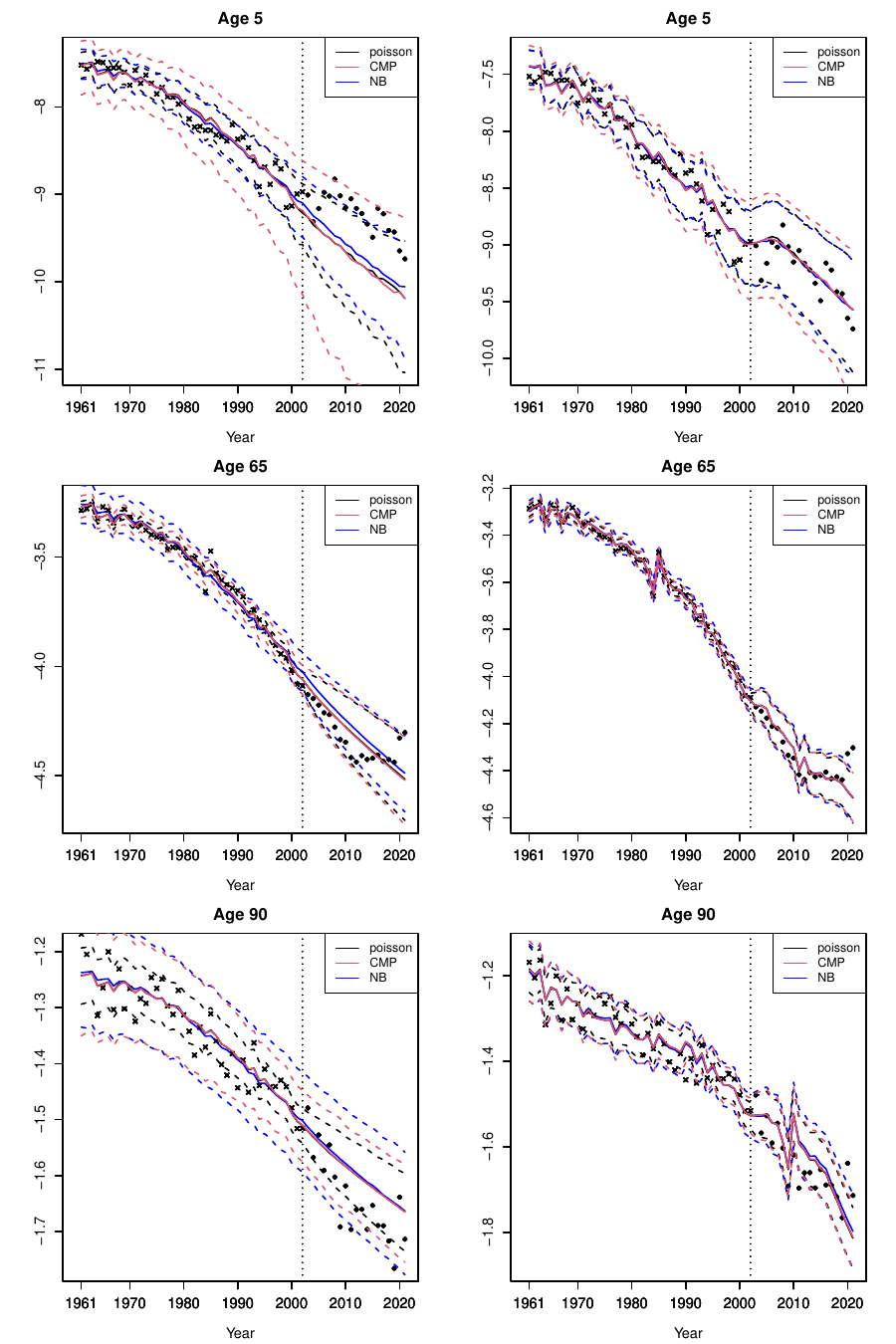}
\caption{\label{fig:projection_logrates_age5_age65_age90}Plots of observed and holdout log crude death rates, fitted log crude death rates and the associated $19$-year ahead projection of the crude log death rates for several chosen ages under the LC (left panels) and LCC (right panels) rate models, accompanied by $95\%$ credible intervals in dashed lines.}
\end{figure}

Figure \ref{fig:projection_logrates_age5_age65_age90} shows the fitted and projected crude death rates for several chosen ages when $h=19$. Also included are the observed and holdout crude mortality rates. Generally speaking, both the CMP and NB specifications yield mortality forecasts with good properties, with close adherence to the holdout crude rates and 95\% credible intervals that have high coverage rates. By contrast, the Poisson specification yields credible intervals that are overly narrow, highlighting the importance of accounting for overdispersion. Comparing between the CMP and NB specifications, the CMP specification produces wider 95\% credible intervals for earlier ages, while the reverse is true for the NB case where it generates wider intervals for older ages. 

Incorporating cohort effects generally leads to a better description  of the underlying mortality trend for both fitted and projected rates. In particular, the $95\%$ prediction intervals of the cohort models (all specifications) offer realistic coverage rates for including the holdout rates, despite having narrower widths. Models without cohort effects generally fail to capture (and hence project) the mortality trends, leading to unrealistic median projections and unnecessarily wide intervals. Simultaneous inclusion of overdispersion and cohort components facilitates the calibration between signals and errors, reducing the bias in the mortality projection with adequately wide intervals to characterise uncertainty. This is consistent with the findings of \cite{wong_2023}. Comparing the various specifications, again the Poisson specification gives low coverages due to overly narrow intervals, while CMP and NB specifications yield sufficiently wide intervals with high coverage. Note that the two most recent mortality rates were observed to be unusually high due to Covid-19 effects, which we do not address here.


\subsection{Sensitivity analysis}

\begin{figure}[htbp]
\centering\includegraphics[scale=0.7]{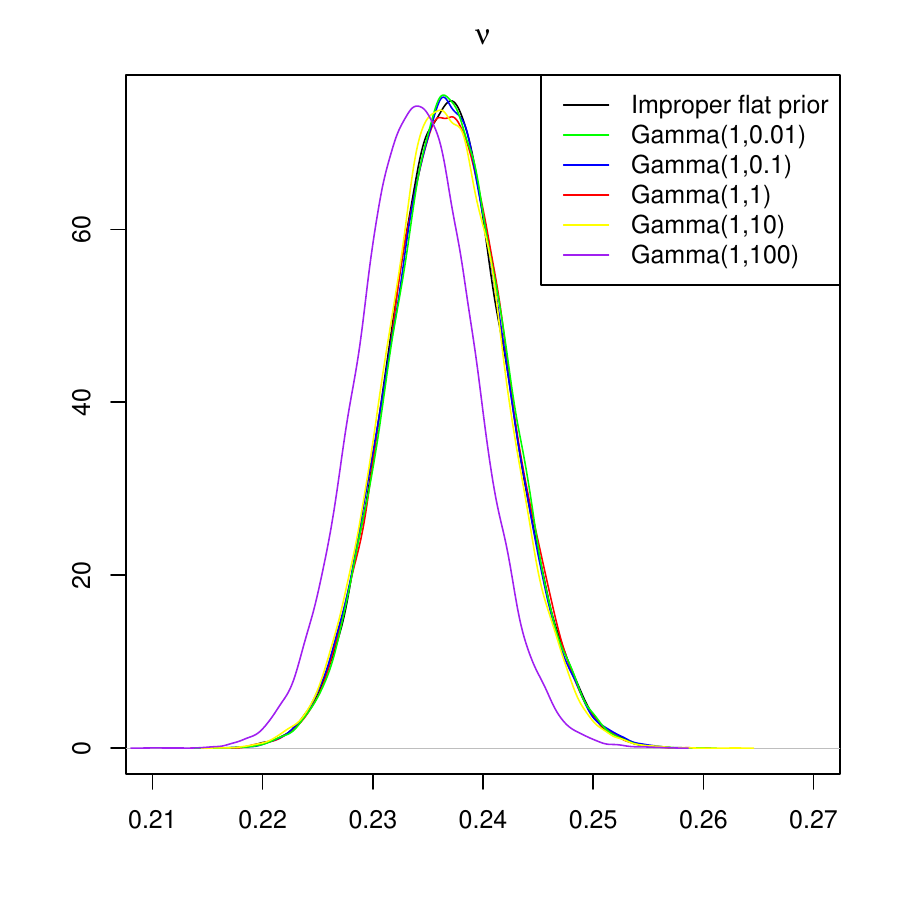}
\caption{Kernel density plots of the posteriors of $\nu$ for the CMP--LC model under various prior specifications.}
\label{fig:CMP_LC_sensitivity_male}
\end{figure}

Finally, we examine the influence of prior specification on the estimation of $\nu$ and assess the sensitivity of the results to the hyperparameters $a$ and $b$. For illustration, this analysis is performed on the CMP--LC model only. Figure \ref{fig:CMP_LC_sensitivity_male} displays the posterior distributions of $\nu$ under various Gamma prior specifications, as described in Section~\ref{sec:prior}.

We first observe that the posterior distributions of $\nu$ under the priors $\text{Gamma}(1, 0.01)$, $\text{Gamma}(1, 0.1)$, $\text{Gamma}(1, 1)$, and $\text{Gamma}(1, 10)$ closely resemble that obtained using a flat prior. This is unsurprisingly expected, as these priors have variances substantially larger than the posterior variance of $\nu$ (approximately $3 \times 10^{-5}$), and therefore exert minimal influence on the posterior estimation. In these cases, the estimation of $\nu$ is largely driven by the data.

An exception arises when the prior $\nu \sim \text{Gamma}(1,100)$ is specified. This prior has a mean of $0.01$ and a variance of $1 \times 10^{-4}$, which is of comparable magnitude to the variance implied by the data. As a result, the prior is sufficiently informative to influence the Bayesian learning process, shifting the posterior distribution of $\nu$ slightly towards the prior mean of $0.01$. This reflects the increased certainty assigned to the prior, relative to other specifications, resulting in a more pronounced impact on the Bayesian estimation.

In summary, as the variance of the Gamma prior decreases (i.\,e., for larger values of $b$), the prior contributes a more prominent role in the estimation of $\nu$.

\section{Conclusion}
\label{sec:conclusion}

In this paper, we introduce the CMP distribution as a flexible framework for modelling death counts. The distribution is embedded within the LC and LCC mortality rate models and estimated using a Bayesian approach. The proposed framework highlights the ability of the CMP  distribution to accommodate underdispersion, equidispersion, and overdispersion through an explicit dispersion parameter, $\nu$, whose value can be inferred directly from the data. By doing so, this approach allows all relevant sources of uncertainty to be coherently incorporated, particularly in settings where overdispersion is present. To facilitate Bayesian inference, we adopt a non-informative gamma prior for the dispersion parameter and develop a corresponding MCMC algorithm to generate posterior samples. 

The proposed methodology is illustrated using England and Wales male mortality data across the entire age range. Empirical results demonstrate that the CMP-based specification led to superior in-sample and out-of-sample performances, as assessed through residual diagnostics and a range of proper scoring rules, when compared with commonly used alternative distributions, such as the Poisson and NB models. These findings underscore the importance of combining flexible distributional assumptions with appropriate structural components in mortality modelling. In particular, incorporating cohort effects alongside a more flexible distribution that accommodates dispersion leads to improved calibration between systematic mortality trends and random variation. A sensitivity analysis was also conducted to assess the impact of prior choices on Bayesian estimation, and the results indicate under the proposed specification appears robust.  

Several additional avenues for future research emerge from our work. First, the time-series dynamics of the period and cohort effects, $\kappa_t$ and $\gamma_c$, were specified using standard formulations from the literature. Extending this framework to allow for formal model selection among alternative time-series specifications may further enhance model stability, particularly when applied to other mortality datasets, although the gains may be modest for the data considered here. Second, the dispersion parameter $\nu$ was assumed to be constant across ages and periods, capturing only a single overall level of dispersion. This may be considered restrictive, as dispersion could vary systematically across the mortality surface. Allowing for age-specific or time-specific dispersion parameters, such as $\nu_x$ or $\nu_t$, represents a promising direction for future research.

Overall, our study demonstrates that integrating the CMP distribution into established mortality models provides a powerful and flexible approach for analysing death counts. By explicitly modelling dispersion within a coherent Bayesian framework, the proposed methodology delivers improved model fit, richer inference, and a solid foundation for further methodological and applied work in mortality modelling and projections.

\section*{Acknowledgment}
The authors greatly appreciate the comments from the editor, associate editor, and all reviewers involved, who helped improving the quality of the paper. 

\section*{Funding}
This research received no specific grant from any funding agency in the public, commercial, or not-for-profit sectors.

\section*{Conflict of Interests}
The authors declare that they have no conflicts of interest regarding the publication of this article. No financial or personal relationships that could have influenced the work have been disclosed.

\section*{Data Availability Statement}
The data and code that support the findings of this study are available from the following GitHub repository [to be included if accepted].

\appendix
\setcounter{secnumdepth}{0}
\section{Appendix: Supplementary Material}
Supplementary material related to this research may be found in the online version of the article at the publisher's website.

\bibliographystyle{chicago}
\bibliography{reference}

\end{document}